\documentstyle[epsf,twocolumn,seceq]{orb2001}

\title
{
Critical Behaviour near the Mott Transition \\
in a Two-Band Hubbard Model
}

\def\runtitle
{
Critical behaviour near the Mott transition \\
in a two-Band Hubbard model
}

\author
{Yasuo {\sc{Ohashi}}\footnote{E-mail: ohashi@edu2.phys.nagoya-u.ac.jp} and Yoshiaki {\sc{\=Ono}}\footnote{E-mail: c42545a@nucc.cc.nagoya-u.ac.jp}
}

\def\runauthor
{
Yasuo {\sc{Ohashi}} and Yoshiaki {\sc{\=Ono}}
}

\inst
{
Department of Physics, Nagoya University, Furo-cho, Chikusa-ku, Nagoya 464-8602, Japan\\
}

\abst
{
The Mott metal-insulator transition in the two-band Hubbard model in infinite dimensions is studied by using the linearized dynamical mean-field theory. The discontinuity in the chemical potential for the change from hole to electron doping is calculated analytically  as a function of the on-site Coulomb interaction $U$, and the charge-transfer energy $\Delta$ between the $d$- and $p$-orbitals, transfer integrals $t_{pd}$, $t_{pp}$, $t_{dd}$ between $p$-$d$, $p$-$p$ and $d$-$d$ sites respectively. The critical behaviour of the quasiparticle weight is also obtained analytically. 
}

\kword
{
Mott transition, metal$-$insulator transition, two-band Hubbard model, dynamical mean-field theory, infinite dimensions
}

\def\nn{\nonumber}

\begin{document}
\sloppy
\maketitle

\section{Introduction}

The Mott metal$-$insulator transition (MIT) is observed in various transition metal oxides and have been extensively studied for years.
The dynamical mean-field theory (DMFT)\cite{Georges1, Metzner}, which becomes exact in the infinite spatial dimensions, has been used in various models and have given plausible results for the MIT. 
However, it is difficult to obtain the critical behaviour of the MIT numerically as the quasiparticle peak becomes extremely sharp in the limit of the MIT. 
There is an alternative way to resolve the adifficulty. The linearized DMFT (LDMF) \cite{Bulla, Ohashi, Ono, Ono3} is an approximate but analytical method which becomes the most correct on the MIT point. 

There are two types of the MIT in real materials: the Mott-Hubbard (MH) type in which the energy gap is given roughly by the Coulomb interaction and the charge-transfer (CT) type in which the energy gap is given roughly by the charge transfer energy. In order to describe the two types of the MIT, we need at least the two-band Hubbard model which is characterized by the Coulomb interaction $U$, charge transfer energy $\Delta$ and $p$-$d$, $p$-$p$ and $d$-$d$ transfer integrals $t_{pd}$, $t_{pp}$ and $t_{dd}$ \cite{Ono}. 

The critical behaviour of the MIT has been studied using the LDMF for the single-band Hubbard model \cite{Bulla, Ono3}. It has also been studied for the two-band Hubbard model in our previous paper \cite{Ohashi}. 
However, effects of $t_{pp}$ and $t_{dd}$ were not considered there, which are not negligible in actual compounds but make important contribution to the critical behaviour near the MIT, as found to be significant to determine the phase boundary of the MIT \cite{Ono}. In the present paper, we study the discontinuity in the chemical potential and the critical behaviour of the MIT in the two-band Hubbard model in the presence of $t_{pp}$ and $t_{dd}$. 


\section{Formulation}

We consider the two-band Hubbard model \cite{Ono}, 
\begin{eqnarray} 
H &=& \frac{t_{pd}}{\sqrt{q}} \sum_{\langle i,j \rangle,\sigma}
        ( d_{i\sigma}^{\dagger} p_{j\sigma} + h.c.)
      + U \sum_{i} d^{\dagger}_{i\uparrow}d_{i\uparrow} 
                   d^{\dagger}_{i\downarrow}d_{i\downarrow}  
        \nonumber \\
  &&\hspace{-4mm}{}+  \frac{t_{dd}}{q} \sum_{\langle i,i' \rangle ,\sigma} 
       (d_{i\sigma}^{\dagger}d_{i'\sigma} + h.c.)
     + \epsilon_{d0} \sum_{i,\sigma} d_{i\sigma}^{\dagger} d_{i\sigma}
 \nonumber \\
  &&\hspace{-4mm}{}+  \frac{t_{pp}}{q} \sum_{\langle j,j' \rangle ,\sigma} 
       (p_{j\sigma}^{\dagger}p_{j'\sigma} + h.c.)
     + \epsilon_{p0} \sum_{j,\sigma} p_{j\sigma}^{\dagger} p_{j\sigma},
        \label{DP}
\end{eqnarray} 
where $d^{\dagger}_{i\sigma}$ and $p_{j\sigma}^{\dagger}$ are creation operators for an electron with spin $\sigma$ in the $d$-orbital at site $i$ and in the $p$-orbital at site $j$, respectively. 
$t_{pd}$, $t_{dd}$ and $t_{pp}$ are the transfer integrals between the nearest neighbor $p$-$d$, $p$-$p$ and $d$-$d$ orbitals, respectively. In eq.~(\ref{DP}), we assume that $p$- and $d$-orbitals are on different sub-lattices of a bipartite lattice, more explicitly, a Bethe lattice with connectivity $q$. 
In the limit $q = \infty$, the self-consistency equations for the local Green's functions are given by \cite{Georges1}
\begin{eqnarray} 
{\cal G}_0(z)^{-1}&=& z -\epsilon_{d} -t_{pd}^2 G_p(z) -t_{dd}^2 G_d(z), \label{SCE3A} \\
G_p(z)^{-1}&=& z  - \epsilon_{p} - t_{pd}^2 G_d(z) -t_{pp}^2 G_p(z), \label{SCE3B}
\end{eqnarray} 
where $G_p(z)$ is the local Green's function for the $p$-electron and $G_d(z)$ is that for the $d$-electron; 
$\epsilon_{d}\equiv \epsilon_{d0}-\mu=-\mu$ and 
$\epsilon_{p}\equiv \epsilon_{p0}-\mu=\Delta-\mu$, where we set $\epsilon_{d0}=0$ and the CT energy $\Delta$ is defined by $\Delta \equiv \epsilon_{p}-\epsilon_{d}>0$.

In the LDMF, the two-band Hubbard model eq.~(\ref{DP}) is mapped onto a two-site Anderson model
\begin{eqnarray}
H_{\rm 2-site}&=& \epsilon_f \sum_{\sigma} f^{\dagger}_\sigma f_\sigma +U f^{\dagger}_{\uparrow}f_{\uparrow}f^{\dagger}_{\downarrow}f_{\downarrow} \nonumber \\ 
&&{}+ \epsilon_c \sum_{\sigma} c^{\dagger}_{\sigma} c_{\sigma} + V \sum_{\sigma}(f^{\dagger}_{\sigma}c_{\sigma}+c^{\dagger}_{\sigma}f_{\sigma}),\label{2-site}
\end{eqnarray}
with $\epsilon_c=0$ and $\epsilon_f=\epsilon_d=-\mu$. 
In the model eq.~(\ref{2-site}), the non-interacting impurity Green's function is: 
$
{\cal G}_0(z)^{-1} = z-\epsilon_f- \frac{V_{\rm }^2}{z}. \label{G0}
$
In the limit $V\to 0$, the local Green's functions are given by 
$G_d(z) = \frac{Z_d}{z}$ and 
$G_p(z) = \frac{Z_p}{z},$\cite{Ono, Ohashi}
near the Fermi level with small weights $Z_d \to 0$ and $Z_p \to 0$. Then the self-consistency equations (\ref{SCE3A}) and (\ref{SCE3B}) are reduced to a simple equation 
\begin{equation}
t_{pd}^2 Z_p + t_{dd}^2 Z_d=V_{\rm }^2. \label{SCE4} 
\end{equation}
To second order in $V$, the quasiparticle weights for the $d$-electron and the $p$-electron are given by \cite{FandA}
\begin{eqnarray}
Z_d&=& F(U,\mu) V^2 ,\label{Z_d}\\ 
Z_p&=& A(t_{pd},t_{pp},U,\Delta,\mu)V^2, \label{Z_p}
\end{eqnarray}
where
\begin{eqnarray}
&&\hspace*{0mm}F(U,\mu)	=  \frac{5}{2\mu^2} 
             + \frac{5}{2(U-\mu)^2}
             + \frac{4}{\mu(U-\mu)} , \label{F}\\
&&\hspace*{0mm}A(t_{pd},t_{pp},U,\Delta,\mu)	= \frac{t_{pd}^2 F(U,\mu)}{
      E_p^2-t_{pp}^2 }, \label{A}
\end{eqnarray}
and
\begin{eqnarray}
\hspace*{-8mm}	E_p &=& \Delta-\mu -   \frac{t_{pp}^2}{{\Delta-\mu}}-t_{pd}^2\left\{-\frac{1}{2\mu}+\frac{1}{2(U-\mu)}\right\}  .
\end{eqnarray}
Substituting eqs.(\ref{Z_d}) and (\ref{Z_p}) with eqs.(\ref{F}) and (\ref{A}) into eq.~(\ref{SCE4}) we obtain an equation to determine the MIT point,
\begin{eqnarray}
P(t_{pd},t_{pp},t_{dd},U,\Delta,\mu)=1,\label{P1}
\end{eqnarray}
where
\begin{eqnarray}
P&=&t_{pd}^2A(t_{pd},t_{pp},U,\Delta,\mu)+t_{dd}^2 F(U,\mu).
\end{eqnarray}
After self-consistency cycle, $V_{\rm }$ increase exponentially with iteration number for $P>1$, and then the single pole approximation breaks down resulting in the metallic solution, on the other hand $P<1$ correspond to the insulating solution. 
In eq.~(\ref{P1}), $P$ includes the chemical potential $\mu$ which has to be determined explicitly to obtain the critical values of the MIT. We can use a certain condition to determine $\mu$, based on the fact that at the MIT point $P$ has a minimum value as a function of $\mu$\cite{Ohashi}. This condition gives 
\begin{eqnarray} 
\frac{\partial}{\partial \mu}
    P(t_{pd},t_{pp},t_{dd},U,\Delta,\mu) = 0. 
\label{P2}
\end{eqnarray}

\section{Discontinuity in the Chemical Potential}

\begin{figure}[b]
\vspace{5mm}
\begin{center}
\leavevmode
\epsfxsize=8cm
    \epsffile{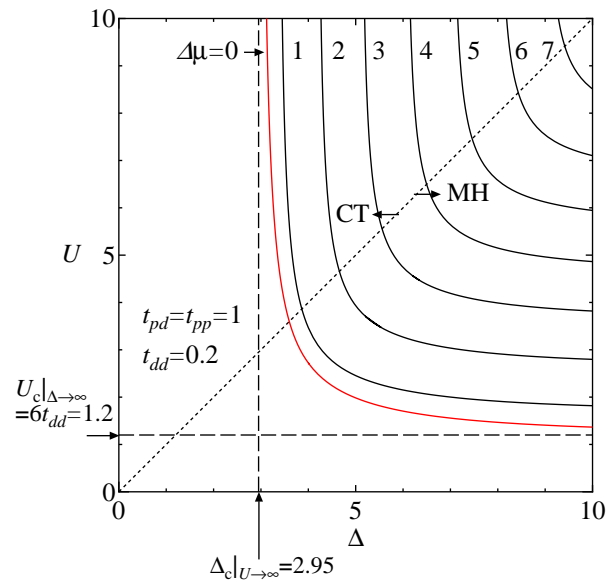}
\caption{The contour map for the discontinuity in the chemical potential ${\it \Delta}\mu=0,1,2,3,4,5,6,7$ at $t_{pd}=t_{pp}=1$ and $t_{dd}=0.2$. The dotted line represents a boundary separating the MH and CT regime.
}
\label{Contour}
\end{center}
\end{figure}

By solving eqs.~(\ref{P1}) and (\ref{P2}), the MIT phase boundary is determined and shown in Fig.~\ref{Contour} (see ${\it \Delta} \mu=0$). In the insulating regime, the chemical potential $\mu_{\pm}$ in the limit $n\to 1\pm0_+$ is obtained from eq.~(\ref{P1}): $P|_{\mu=\mu_\pm}=1$ \cite{Ohashi}, which yields the discontinuity in the chemical potential, ${\it \Delta}\mu=\mu_+ -\mu_-$. 
In Fig.~\ref{Contour}, we show the contour map for  ${\it \Delta}\mu$ on the $\Delta$-$U$ plane, where the diagonal dotted line represents a boundary separating the CT and the MH regimes, and the dashed lines are the critical values of $U_{\rm c}$ and $\Delta_{\rm c}$ in the limit of $\Delta\to\infty$ and $U\to\infty$, respectively. The discontinuity in the chemical potential smoothly connects the Mott-Hubbard type and the charge-transfer type insulators. 
In the limit  $\Delta\to\infty$, the critical value of $U_{\rm c}=6\ t_{dd}$ is the same as that for the corresponding single-band Hubbard model. \cite{Bulla, Ono3}
%

%
%

\section{Critical Behaviour near the Mott MIT}

We discuss the critical behaviour in the vicinity of the Mott MIT at half-filling in the two-band Hubbard model. In this case, we need the result of the quasiparticle weight for the $d$-electron up to fourth order in $V$ 
, 
\begin{eqnarray} 
Z_d=F(U,\mu)V^2-G(U,\mu)V^4, \label{Z_d2}
\end{eqnarray} 
where $F$ is given in eq.~(\ref{F}) and 
\begin{eqnarray} 
G(U,\mu)&=&\frac{29}{2\mu^4}+\frac{24}{\mu^3(U-\mu)}+\frac{22}{\mu^2(U-\mu)^2}
              \nonumber \\
        & &    +\frac{24}{\mu(U-\mu)^3}+\frac{29}{2(U-\mu)^4}. 
\end{eqnarray} 
The quasiparticle weight for the $p$-electron $Z_p$ is also calculated up to fourth order in $V$, 
\begin{eqnarray}
\hspace{-7mm}Z_p \hspace{-1mm}=\hspace{-1mm} A(t_{pd},t_{pp},U,\Delta,\mu)V^2 \hspace{-1mm}-\hspace{-1mm} B(t_{pd},t_{pp},U,\Delta,\mu)V^4 , 
	\label{Z_p2}
\end{eqnarray}
where $A$ is given in eq.~(\ref{A}) and
\begin{eqnarray}
B(t_{pd},t_{pp},U,\Delta,\mu)  
&=&t_{pd}^2\frac{G}{E_p^2-t_{pp}^2} 
+t_{pd}^4  \frac{2 F \,B_1 E_p}{(E_p^2-t_{pp}^2)^2}\nn\\
&&\hspace{-15mm}{}+t_{pd}^4 \frac{F^2\big\{2 t_{pp}^2 +3B_2E_p (\Delta-\mu)\big\}}{(E_p^2-t_{pp}^2)^3 (\Delta-\mu)}
,  \label{B}
\end{eqnarray}
with
\begin{eqnarray}
\hspace{-5mm}    B_1 &=&  \frac{-3\,U^3 + 7\,U^2\,{\mu} - 
      3\,U\,{{\mu}}^2 + 
      2\,{{\mu}}^3}{{{\mu}}^3\,{\left( U - {\mu} \right)
          }^3} , \\
\hspace{-5mm}	B_2 &=& 1+t_{pd}^2\left\{\frac{1}{2\mu^2}+\frac{1}{2(U-\mu)^2}\right\}+ \frac{{{t_{pp}}}^2}
   {(\Delta-\mu)^2} .
\end{eqnarray}
Substituting $V^2$ from the self-consistency equation (\ref{SCE4}) into eqs.~(\ref{Z_d2}) and (\ref{Z_p2}), we obtain 
\begin{eqnarray}
  Z_d^{{\rm }} &=& F(U,\mu)
  \frac{P-1}{Q},\label{Z_d3}
\end{eqnarray}
where
\begin{eqnarray}
Q&=&t_{pd}^2B(t_{pd},t_{pp},U,\Delta,\mu)+t_{dd}^2 G(U,\mu).
\end{eqnarray}

\begin{figure}[b]
\vspace{5mm}
\begin{center}
\leavevmode
\epsfxsize=8cm
   \epsffile{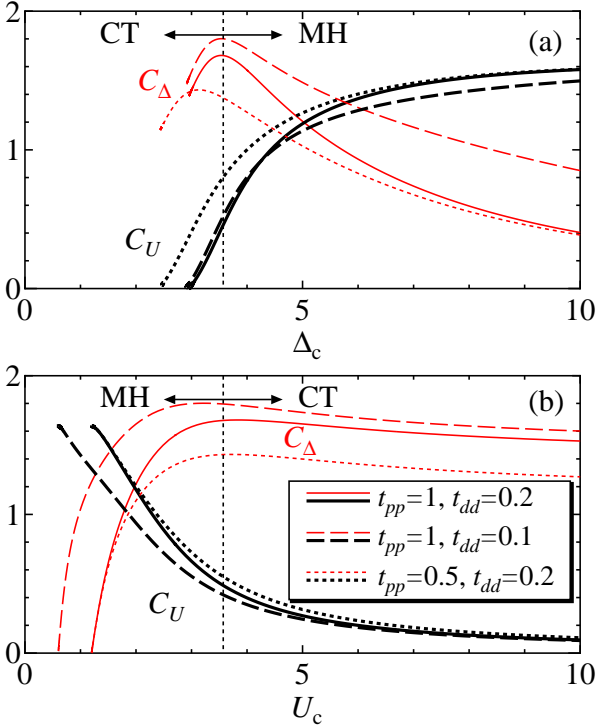}
\caption{
The coefficients $C_{\Delta}$ (thin lines) and $C_{U}$ (thick lines) of the quasiparticle weights for the $d$-electrons 
$
Z_{d}= C_{\Delta}\left(1-\frac{\Delta}{\Delta_{\rm c}}\right) 
$
and 
$
Z_{d} = C_{U} \left(1-\frac{U}{U_{\rm c}}\right), 
$
respectively, as functions of $\Delta_{\rm c}$ (a) and $U_{\rm c}$ (b) for several values of $t_{pp}$ and $t_{dd}$ with $t_{pd}=1$. 
The thin vertical dotted line shows a boundary of MH and CT for $
t_{pp}=1$ and $t_{dd}=0.2$.}
\label{alpha}
\end{center}
\end{figure}

At the MIT point with the critical values $U_{\rm c}$, $\Delta_{\rm c}$ and $\mu_{\rm c}$, eqs.~(\ref{Z_d3}) yield $Z_d=0$ from eq.~(\ref{P1}).
When $\Delta$ or $U$ decreases from the MIT point, 
the quasiparticle weight for the $d$-electron $Z_d$ eq.~(\ref{Z_d3}) increases as
\begin{eqnarray}
  Z_d &=& C_{\Delta} \left(1-\frac{\Delta}{\Delta_{\rm c}}\right), \ \ \
\label{ZdD} \\
  Z_d&=& C_{U} \left(1-\frac{U}{U_{\rm c}}\right),  \ \ \ 
\label{ZdU} 
\end{eqnarray}
near the MIT point at half-filling, respectively.
The coefficients $C_\Delta$ and $C_{U}$ are shown in Figs. \ref{alpha}(a) and (b).

When $t_{pp}$ increases, the $p$-component of the quasiparticle weight increases, then, $C_\Delta$ increases and $C_U$ decreases (compare the dotted line with the solid line in Figs. \ref{alpha}(a) and (b)). On the other hand, when $t_{dd}$ increases, the $d$-component of the quasiparticle weight increases, then, $C_\Delta$ decreases and $C_U$ increases (compare the dashed line with the solid line in Figs. \ref{alpha}(a) and (b)).

In the CT regime ($U>\Delta$), the MIT occurs at $\Delta=\Delta_{\rm c}$ when $\Delta$ is varied for a fixed $U$. As $\Delta$ decreases below $\Delta_{\rm c}$ for a fixed $U$, the quasiparticle weight increases as given in eq.~(\ref{ZdD}). 
With increasing $U_{\rm c}$ (decreasing $\Delta_{\rm c}$), the coefficients both $C_\Delta$ and $C_U$ decrease due to the increasing correlation effect as seen in Figs.\ref{alpha}(a) and (b). 
In the limit $U_{\rm c}\to\infty$, the coefficients are 
\begin{eqnarray}
C_{\Delta} = 1.31, \hspace{5mm} C_{U}=0,
\end{eqnarray}
for $t_{pd}=t_{pp}=1$ and $t_{dd}=0.2$.

In the MH regime ($U<\Delta$), the MIT occurs at $U=U_{\rm c}$ when $U$ is varied for a fixed $\Delta$. As $U$ decreases below $U_{\rm c}$ for a fixed $\Delta$, the quasiparticle weight increases as given in eq.~(\ref{ZdU}). 
With increasing $\Delta_{\rm c}$ (decreasing $U_{\rm c}$), $C_U$ increases due to the decreasing correlation effect, while $C_{\Delta}$ decrease due to the decreasing $p$-component of the quasiparticle weight (See Figs.\ref{alpha}(a) and (b)). 
In the limit $\Delta_{\rm c}\to\infty$, the coefficients are 
\begin{eqnarray}
C_{\Delta}=0, \hspace{5mm} C_{U}=\frac{18}{11}, \label{C_Dinf}
\end{eqnarray}
for $t_{pd}=t_{pp}=1$ and $t_{dd}=0.2$. 
In the limit $\Delta_{\rm c}\to\infty$, the effect of the $p$-band becomes irrelevant and, then, $C_\Delta=0$ and $C_U=\frac{18}{11}$ is the same value as that for the single-band Hubbard model \cite{Bulla}. 
We note that, even in this limit, $C_\Delta$ is finite and $C_U \ne \frac{18}{11}$ in the case with $t_{dd}=0$, because the electron has to transfer between $d$ and $p$ orbitals through the hopping integral $t_{pd}$ and the effect of the $p$-band is still relevant as shown in Ref. 4.

\section{Discussion}

In the real materials, the ratio of the transfer integrals are estimated as: ${\rm La_2CuO_4}$, $t_{pd}:t_{pp}=1:0.50$ \cite{Hybertsen}, ${\rm LaTiO_3}$, $t_{pd}:t_{pp}=1:0.15$, ${\rm LaNiO_3}$, $t_{pd}:t_{pp}=1:0.27$ \cite{Slater, Mahadevan} (no data for $t_{dd}$). There is little data for the critical behaviour of the MIT when $U$ or $\Delta$ is varied at the half-filling because of the difficulty of the experiments \cite{Imada}. 
Further progress in both theoretical and experimental study for the critical behaviour of the MIT is expected.





\begin{thebibliography}{}
\bibitem{FandA}
These fanctions $F$ and $A$ are obtained by following the same calculation as shown in the appendix in Ref. 4, where we approximated the $d$- and $p$-Green's function of the right hand side of eqs.~(\ref{SCE3A}) and (\ref{SCE3B}) as three-pole and two-pole functions, respectively.
\end{thebibliography}
\end{document}